\documentclass[aps,twocolumn,prd,superscriptaddress,preprintnumbers,showpacs]{revtex4}
\usepackage{graphicx}

\newcommand{\be}{\begin{equation}}
\newcommand{\ee}{\end{equation}}
\newcommand{\bea}{\begin{eqnarray}}
\newcommand{\eea}{\end{eqnarray}}
\newcommand{\mpl}{M_{\rm Pl}}

\begin{document}

\preprint{}

\title{Super-inflation in Loop Quantum Cosmology }

\author{E. J. Copeland}
\email[]{Ed.Copeland@nottingham.ac.uk}
\affiliation{School of Physics and Astronomy, University of Nottingham, University Park, Nottingham NG7 2RD, UK}
\author{D. J. Mulryne}
\email[]{D.Mulryne@damtp.cam.ac.uk}
\affiliation{Department of Applied Mathematics and Theoretical Physics, Wilberforce Road, Cambridge, CB3 0WA, UK}
\author{N. J. Nunes}
\email[]{nunes@damtp.cam.ac.uk}
\affiliation{Department of Applied Mathematics and Theoretical Physics, Wilberforce Road, Cambridge, CB3 0WA, UK}
\author{M. Shaeri}
\email[]{ppxms1@nottingham.ac.uk }
\affiliation{School of Physics and Astronomy, University of Nottingham, University Park, Nottingham NG7 2RD, UK}

\date{\today}

\begin{abstract}
We investigate the dynamics of super-inflation in two versions of Loop Quantum Cosmology, one in which the Friedmann equation is modified by the presence of inverse volume corrections, and one in which quadratic corrections are important.
Computing the tilt of the power spectrum of the perturbed scalar field in terms
 of fast-roll parameters, we conclude that the first case leads to a power
 spectrum that is scale invariant for steep power law negative potentials and for the second case, scale invariance is obtained for positive potentials that asymptote to a constant value for large values of the scalar field. It is found that in both cases, the horizon problem is solved with only a few $e$-folds of super-inflationary evolution. 
\end{abstract}

%\keywords{}
\maketitle

%%%%%%%%%%%%%%%%%%%%%%%%
\section{Introduction}
%%%%%%%%%%%%%%%%%%%%%%%%
An inflationary epoch 
is currently the most promising model for the
origin of large-scale structure in the universe \cite{inflation}.  The predictions of
inflation are fully compatible with the most recent observations
suggesting that structure
originated from a pattern of near scale-invariant, Gaussian and
adiabatic primordial density fluctuations \cite{Spergel:2006hy}.  
Despite these successes, however, a number of important questions
remain.  In particular, in what fundamental theory will inflation be seen to arise? Having asked the question, it is worth noting that there have been successful implementations of inflation both in the context of ordinary field theory \cite{Copeland:1994vg,Lyth:1998xn} as well as in the context of string and M-theory \cite{Copeland:1994vg,Dvali:1998pa,Burgess:2001fx,Jones:2002cv,Kachru:2003aw,Kachru:2003sx,Firouzjahi:2003zy,Dasgupta:2004dw,BlancoPillado:2004ns,Conlon:2005jm,BlancoPillado:2006he}. Given the importance of inflation and
the need to explore all possibilities of  accommodating it  in alternative theories of quantum gravity, in this paper we turn our attention to inflation in the context of Loop Quantum Gravity (LQG) \cite{LQG}.

LQG is  a background independent and 
non-perturbative canonical 
quantisation of general relativity based on Ashtekar variables:
su(2) valued connections
and conjugate triads.  The variables used in
the quantisation scheme are then holonomies of the connection and
fluxes of the triad. The restriction of LQG to symmetric states gives rise to Loop Quantum
Cosmology (LQC) \cite{LQC}. 
Although it is a particular limit of the more general LQG, and therefore can not be said to have generic features, LQC has produced a number of intriguing results and resolved many
problematic issues present in the earlier Wheeler de Witt quantum cosmology.
In particular, LQC can lead to a non-singular quantum evolution \cite{NS}, with the 
origin of the non-singular behaviour being traced to the methods used
to quantise inverse triad operators in LQG \cite{Thiemann:1996aw,inverse}.  

While the consequences of this quantum evolution are fascinating,
it is difficult to connect it to existing theories
of the early universe which tend to be based on classical dynamics, 
and in particular
to inflation. Therefore, an approach to LQC has been developed
in which effective or `semi-classical' equations are derived and
studied.  In the isotropic setting, the effective equations 
which have been studied
predominately to date include high energy
modifications to the classical dynamics which originate 
from the spectra of quantum operators related to 
the inverse scale factor
\cite{inverse,Bojowald:2002ny,Bojowald:2002nz,Vandersloot:2005kh}.
In the context of scalar field driven inflation, a number of
important effects follow from these modifications.  These include the 
possibility that the field can be excited up its self-interaction
potential \cite{Bojowald:2004xq, Tsujikawa:2003vr, Lidsey:2004ef,
Mulryne:2004va, Mulryne:2005ef, Nunes:2005ra}, leading to a subsequent
period of slow-roll inflation. Most intriguing of all, however,  
is the presence of a super-inflationary 
phase which occurs during the early phases of the universe's evolution
irrespective of the form of the field's potential
\cite{Bojowald:2002nz,Bojowald:2003mc}.

More recently, however, a further `semi-classical' modification, which
arises from the use of holonomies as a
basic variable in the quantisation scheme, has been derived in the 
isotropic setting \cite{rho2}.  The modification is remarkably simple
and 
takes the form of an additional negative $\rho^2$ term in the
effective Friedmann equation, which appears in addition to the usual 
positive $\rho$ term.  Such a term has a
number of effects, it forces a collapsing universe to undergo a
non-singular bounce once a critical density is reached, and
immediately after this bounce it also causes the universe to undergo 
a period of super-inflation. It is intriguing to note that such a term
also appears in braneworld models with an extra time-like 
dimension \cite{Shtanov:2002mb}.

While the two sets of modifications discussed above have rather different origins, 
it appears that the qualitative effects of both the inverse scale factor
effects and the $\rho^2$ term are rather similar. In particular, they both give rise to  
a period of super-inflation during which the Hubble factor rapidly increases, rather than
remaining nearly constant as is the case during standard slow-roll inflation.  
Given that such a super-accelerating 
phase appears to be a robust prediction 
of LQC, it is important to study both the background dynamics, and 
particularly the cosmological perturbations, which such a  
phase gives rise to. 
Considering a universe sourced by a scalar field, a number of important results have already 
been obtained. A scaling solution for the effective equations which 
arise from the inverse scale factor modifications has been derived \cite{Lidsey:2004uz}, 
and a number of attempts made at studying perturbations in the
super-inflationary regime \cite{Hossain:2004wm, Mulryne:2006cz}.
Similarly the scaling solution for the $\rho^2$ effective equation 
has also been derived \cite{Copeland:2004qe, Singh:2006sg}.

It is also interesting to note the close
connections  between the super-inflationary phases in LQC and 
the evolution of a universe sourced by a phantom field, 
and with the ekpyrotic
evolution of a collapsing universe \cite{Khoury:2001wf,Khoury:2001bz,Khoury:2001zk,Steinhardt:2001st}.
In all these cases the magnitude of the Hubble
rate grows with time. Moreover, the scale factor duality discussed in
\cite{Lidsey:2004xd} maps the ekpyrotic collapse onto the
super-inflationary scaling solution for the inverse scale factor
modified equations.  
On the other hand, another duality maps the ekpyrotic collapse phase onto the dynamics of a 
universe sourced by a phantom field \cite{Lidsey:2004xd}. These three regimes are
therefore all related to one another. Furthermore, given that the
ekpyrotic collapse is thought to offer a method for the generation 
of scale-invariant perturbations \cite{Gratton:2003pe} (as is the dual super-inflationary
phase sourced by a phantom field \cite{Piao:2003ty}), it is reasonable to expect that a similar
mechanism may operate in the super-inflationary phases of LQC. 
Indeed such a mechanism has been discussed previously \cite{Mulryne:2006cz}, though its
relation to ekpyrotic and phantom models was not emphasised.

In this study we aim to explore further the phenomenology of 
super-inflation in LQC.  One complication, however, is that the
relative status of the two sets of modifications discussed is at
present unclear.   We therefore take a pragmatic 
approach and study the dynamics when each of the
modifications is considered in turn, but not including both sets of
modifications simultaneously, although we believe it should not be too difficult to 
incorporate them both. A further difficulty is that despite considerable progress the understanding of metric perturbations in LQC is at present incomplete \cite{Bojowald:2006zb}.  We therefore restrict our attention to perturbations in the scalar field as a first approximation. This approach allows us to establish a framework for dealing with perturbations in LQC in which metric perturbations can be incorporated as our understanding advances.

The paper is organised as follows. In section \ref{inverse-vol}, we introduce the cosmological evolution equations which arise in LQC including the inverse volume corrections. Solutions are obtained including those showing scaling behaviour, and the primordial spectrum of scalar perturbations is calculated for each solution in terms of `fast-roll' parameters. The stability of these solutions is then discussed. In section \ref{rhosq}, we analyse the dynamics when the modification is induced by a $\rho^2$ correction to the Friedmann equation. Concentrating on the evolution just after the bounce, we demonstrate the existence of the super-inflationary solution, obtain the scaling dynamics of the system, the primordial spectrum of scalar perturbations as well as the stability of the background solutions.  Section \ref{efolds} discusses the way in which super-inflation in LQC can solve the horizon problem with a small number of e-foldings and we conclude in section
\ref{discussion}.

%%%%%%%%%%%%%%%%%%%%%%%%
\section{Effective field equations with LQC inverse volume corrections}
\label{inverse-vol}
%%%%%%%%%%%%%%%%%%%%%%%%
The first set of modified equations which we consider are those which
incorporate two functions,  $D_{j,l}(a)$ and 
$S_{j}(a)$, into the dynamics. These functions arise because of the
presence of powers of the inverse scale factor in the
Hamiltonian constraint for an isotropic and
homogeneous universe.  A full discussion of the origin of these terms
can be found in Ref.~\cite{Bojowald:2002ny,Vandersloot:2005kh} (a summary can be found in appendix B of Ref.~\cite{Magueijo:2007wf}), but here we simply state their basic
properties. We are implicitly considering either positively curved or topologically compact models. This ensures that the size of the fiducial cell does not enter in the equations of motion. $D$ and $S$ are both functions of the scale factor, and their
form changes depending on the values of two ambiguity parameters: $l$
which takes values in the range $0<l<1$, and $j$ which takes half
integer values. When the scale factor approaches zero, $D$ and $S$
also approach zero, whereas as $a$ increases above the critical value
$a_\star$ which depends on $j$, they both tend to unity.

The modified Friedmann equation is given by 
\be
\label{FriedD}
H^2 \equiv \left (\frac{\dot{a}}{a}\right)^2 = \frac{\kappa^2}{3} S \left( \frac{\dot{\phi}^2}{2D}+V(\phi) \right)\,~,
\ee
where a dot denotes differentiation with respect to cosmic time
$t$, and $\kappa^2=8\pi G$. In what follows we choose units in which $\kappa = 1$. 
We have omitted the curvature contribution as we assume either a compact flat universe or that the curvature term rapidly becomes sub-dominant and that it can be safely neglected.
The equation of motion for the scalar field takes the form
\be
\label{ScalarD}
\ddot{\phi} + 3H \left( 1- \frac{1}{3} \frac {d\ln D}{d \ln a} \right) \dot{\phi} + DV_{,\phi} =0\,,
\ee
A subscript $\phi$ means differentiation with respect to the field.
These equations can also be combined to give the Raychaudhuri equation
\be
\label{HdotD}
\dot{H} =  - \frac{S \,\dot{\phi}^2}{2 D}
\left[ 1-\frac{1}{6} \frac{d \ln D}{d \ln a} -\frac{1}{6} \frac{d \ln S}{d
  \ln a}  \right] + \frac{S\,V}{6}\frac{d \ln S}{d
  \ln a } \,~.
\label{raycheq}
\ee

%%%%%%%%%%%%%%%%%%%%%%%%
\subsection{Scaling dynamics}
%%%%%%%%%%%%%%%%%%%%%%%%

We will be interested
in the regime $a \ll a_\star$, where the function $D_{j,l}(a)$ may be
approximated by a power law of the form $D(a) = D_{\star} a^n$, with $D_{\star} =
\left( 3/(3+2l) \right)^{3/2(1-l)} \, a_{\star} ^{3(l-3)/(1-l)}$ and $n = 3(3-l)/(1-l)$
takes values in the range $9<n<\infty$.
Likewise, the function $S(a)$ may be similarly 
approximated by $S(a) = S_\star a^r$, where $S_\star = (3/2) a_\star^{-3}$ 
and $r = 3$, though we keep  $r$ arbitrary in our calculations for generality.
For $a \gg a_\star$ , $S_\star \approx D_\star \approx 1$
and $r = n = 0$.
Inserting this form for the functions $S$ and $D$ into Eq.~(\ref{HdotD}), we can clearly see that for an expanding universe, and
with $n>6+r$ which occurs for all $l$, $\dot{H}$ is necessarily
positive (assuming that the potential is either positive or the term
involving $SV$ can be neglected). 
Hence super-inflation is occurring. We will confine ourselves to
these situations in what follows.

To study this regime further, it proves convenient to introduce the
variables
\be
\label{defx}
x\equiv \frac{\dot{\phi}}{\sqrt{2D\rho}}, \hspace{1cm} y\equiv \frac{\sqrt{|V|}}{\sqrt{\rho}}\,.
\ee
where $\rho \equiv \dot{\phi}^2/2D + V(\phi)$.
Using these definitions, the equation of motion for the scalar field (\ref{ScalarD}) can
be written for an expanding universe in terms of a system of first
order differential equations as
\begin{eqnarray}
\label{xN}
x_{,N} &=& -3 \alpha x \pm \sqrt{\frac{3}{2}} \lambda y^2 + 3 \alpha x^3 \,, \\
\label{yN}
y_{,N} &=& -\sqrt{\frac{3}{2}} \lambda xy + 3 \alpha x^2 y \,, \\
\label{lambdaN}
\lambda_{,N} &=& -\sqrt{6} \lambda^2 (\Gamma-1) x + \frac{1}{2}(n-r)\lambda  \,,
\end{eqnarray}
where
\be
\label{deflambda}
\lambda \equiv -\sqrt{\frac{D}{S}} \, \frac{V_{,\phi}}{V} \,, \hspace{1cm}
\Gamma \equiv \frac{V \, V_{,\phi \phi}}{V_{,\phi}^2} \,,
\ee
with $\alpha = 1-n/6 < 0$ and $N = \ln a$.
These variables are subject to the constraint equation
\be
\label{constraint_1}
x^2 \pm y^2=1.
\ee
The plus and minus signs correspond to positive and negative potentials, respectively.
Using the constraint equation to substitute for $y$ in Eq. (\ref{xN}) renders Eq. (\ref{yN})
redundant.

The resulting system defined by Eqs.~(\ref{xN})
together with the constraint 
equation and  (\ref{lambdaN})  has three fixed points for $\lambda \neq 0$.
Two of them represent kinetic energy dominated solutions,
valid for all values of $\lambda$:
\begin{eqnarray}
\label{FixedPoints1}
x &=& -1 \,, \hspace{0.5cm} y = 0 \,, \hspace{0.5cm} \Gamma = 1 - \frac{\sqrt{6}}{12 \lambda} (n-r)   \,,\\
\label{FixedPoints3}
x &=& +1 \,, \hspace{0.5cm} y = 0 \,, \hspace{0.5cm} \Gamma = 1 + \frac{\sqrt{6}}{12 \lambda} (n-r)  \,,
\end{eqnarray}
and the third point is a scaling solution for which the kinetic and potential energies evolve in a constant ratio to one another:
\begin{eqnarray}
\label{FixedPoints2}
 x &=& \frac{\lambda}{\sqrt{6} \, \alpha}, \hspace{1cm}
 y = \sqrt{\pm\left(1-\frac{\lambda^2}{6\alpha^2}\right)} \,, \\
 \Gamma &=&  1 + \frac{\alpha}{2\lambda ^2} (n-r)  \,.
\end{eqnarray}

The scaling solution is therefore well defined for $\lambda^2 < 6 \alpha^2$ for positive 
potentials and for $\lambda^2 > 6 \alpha^2$ for negative potentials. In the remainder of this 
analysis, we will focus mainly on the scaling solution for negative potentials as this is the case
that, as we shall see, leads to a scale invariant power spectrum of
the perturbed field. For this case one can check that the universe is
undergoing super-inflationary expansion.

Considering the fixed point for the scaling
solution (\ref{FixedPoints2}), one can write
\be
\frac{\dot{\phi}}{\sqrt{2 D \rho}}= \sqrt{\frac{S}{D}} \frac{\phi_{,N}}{\sqrt{6 }} = \frac{\lambda}{\sqrt{6}\alpha} \,,
\ee
which upon integration gives
\be
\label{solphi}
\phi = \frac{2\lambda}{(n-r)\alpha} \sqrt{\frac{D}{S}} \,,
\ee
where we have set the integration constant to zero without loss of generality.

Then inserting this relation into the definition of $\lambda$ in (\ref{deflambda})
gives
\be
\label{Pot1}
V = V_0 \, \phi^\beta \,,
\ee
where $\beta= -2 \lambda^2/(n-r) \alpha > 0$.

Considering now the fixed point for $y$ we have 
\be
\label{fixedy}
\frac{V}{\rho} = \frac{V \, S}{3 H^2} = 1-\frac{\lambda^2}{6 \alpha^2} \,.
\ee
Differentiating Eq.~(\ref{solphi}) and eliminating $\dot{\phi}$
using Eq.~(\ref{FixedPoints2}), then substituting for $\rho$ in terms of
$V$ using Eq.~(\ref{fixedy}) and
finally substituting for $V$ in terms of $a$ using Eq.~(\ref{Pot1})
and Eq.~(\ref{solphi}), one
obtains an expression between $a$ and $\dot{a}$.  Then by integrating this
expression we can determine that the scale factor evolves as a power law
in time while the universe evolves according to the scaling solution.
In terms of conformal time $dt = a d\tau$, we find
\be
\label{atau}
a(\tau) = (-\tau)^p \,,
\ee
where for an expanding universe $\tau$ is negative and increasing
towards zero and 
\be
\label{defp}
p = \frac{2\alpha}{2\bar{\epsilon}-(2+r)\alpha} \,,
\ee
where, for direct comparison with previous literature, we have
introduced the slow-roll 
parameter $\bar{\epsilon} \equiv \lambda^2/2$ and $\lambda = -\sqrt{2 \bar{\epsilon}}$ for $\dot{\phi} > 0$.
Using this form of $a$ we find that $H = p/a\tau$, and it is
straightforward to 
show that
\begin{eqnarray}
\label{phiprime}
\phi'(\tau) &=& - \frac{2 \sqrt{2}}{2\bar{\epsilon}-(2+r)\alpha} \sqrt{\frac{D}{S}} \,\frac{1}{\tau} \,, \\
\label{Vtau}
V(\tau) &=& \frac{4(3\alpha^2-\bar{\epsilon})}{(2\bar{\epsilon}-(2+r)\alpha)^2} \, \frac{1}{S (a\tau)^2} \,,
\end{eqnarray}
where a prime means differentiation with respect to conformal time, $\tau$.
Equations (\ref{atau}) - (\ref{Vtau}) form the basis of our analysis. 
This scaling relation (for $S = 1$) was first uncovered in Ref.~\cite{Lidsey:2004uz} using a different procedure.

%%%%%%%%%%%%%%%%%%%%%%%%
\subsection{Power spectrum of the perturbed field}
%%%%%%%%%%%%%%%%%%%%%%%%
For a universe which evolves according to the scaling solution, the
primordial spectrum of scalar perturbations produced by this
super-inflationary phase was previously calculated in Ref.~\cite{Mulryne:2006cz}.
It was found that the spectrum tends to exact scale invariance
for $\beta \gg 1$ (i.e. $\bar{\epsilon} \gg 1$), without any fine tuning
of the quantisation parameters of
LQC. The purpose of this section is to
review how scale invariance arises for the scaling solution
with $\beta \gg 1$, and to generalise the analysis of \cite{Mulryne:2006cz}
in order to allow for potentials which do not give rise to exact scaling
solutions.

In order to calculate the spectrum of perturbations, we
now perturb the scalar field equation.
The perturbation in the field $\delta\phi$
then satisfies the equation
\be
\label{deltaphi}
\delta\phi'' = \left[ -2\frac{a'}{a}+\frac{D_{,\tau}}{D}\right] \, \delta\phi'+
D\left[ \nabla^2 - a^2 V_{,\phi\phi} \right] \delta\phi\,,
\ee
which can be written in the form \cite{Mulryne:2006cz}
\be
\label{u''}
u''+\left( - D\nabla^2+m_{\rm eff}^2\right) \, u = 0 \,,
\ee
where $u$ is defined as $u=aD^{-1/2}\delta\phi$ and the effective mass of the field $u$
is given by
\be
\label{meff}
m_{\rm eff}^2 = -\frac{(aD^{-1/2})''}{aD^{-1/2}}+a^2DV_{,\phi\phi}\,~.
\ee
Decomposing $u$ in Fourier modes $w_k$, that satisfy
\be
\label{w''}
w_k''+\left( - D k^2+m_{\rm eff}^2\right) \, w_k = 0 \,,
\ee
the power spectrum is then given by
\be
{\cal P}_u = \frac{k^3}{2\pi^2} |w_k|^2 \,.
\ee

It was shown in Ref.~\cite{Mulryne:2006cz} that the general solution to Eq.~(\ref{w''}) is
\begin{eqnarray}
\label{w_knormalised}
w_{k}(\tau) &=& \sqrt{\frac{\pi}{2|2+np|}} \,  \left( d_{1}\sqrt{-\tau} \,
H_{|\nu|}^{(1)}(x) \right.\nonumber \\
&~& + \left. d_{2} \sqrt{-\tau} \,H_{|\nu|}^{(2)}(x) \right) \,
\end{eqnarray}
whenever $m_{\rm eff} \tau$ is constant, where
\begin{equation}
\label{nudef}
\nu = - \frac{\sqrt{1-4 \, m_{\rm eff}^2 \, \tau^2}}{2+np}  \,,
\end{equation}
and $d_{1}$ and $d_{2}$ are constants subject to the condition
$|d_{1}|^2-|d_{2}|^2=1$ and $H_{|\nu|}^{(1)}(x)$ and
$H_{|\nu|}^{(2)}(x)$ 
are Hankel functions of the first and second kind, respectively.
In the long wavelength limit, the power spectrum yields
\begin{equation}
\label{Pu}
{\cal P}_{u} \propto  k^{3-2|\nu|} \, (-\tau)^{1-|\nu|(np+2)} \,.
\end{equation}
Scale invariance of the power spectrum is then attained when the spectral tilt 
${\Delta n}_u \equiv 3- 2|\nu| $ is zero. Since for a universe evolving 
according to the scaling solution Eqn.~(\ref{atau}) 
\be
m_{\rm eff}^2 \tau^2 = -2 + (3-2n)p +\frac{1}{2} (6+2n-n^2)p^2 \,,
\ee
we can see from Eq.~(\ref{nudef}) that scale invariance occurs whenever $p\to 0$,
which, as we referred to, does indeed imply that $\bar{\epsilon} \gg 1$ and consequently $V< 0$ from Eq.~(\ref{Vtau}). 
There is one other value of $p$ for which scale invariance is attained, $p=-4/(n+4)$, 
however, we will not consider it any further.

We would now like to generalise the form of the potential we are
dealing with so that it no longer has to be of exactly the form which
gives rise to the scaling solution.
For standard slow-roll inflation, where the kinetic energy is small
compared with the potential energy, it is possible to account for 
potentials of a form more general than a scaling potential 
by introducing slow-roll parameters.  Such parameters parametrise 
the steepness of the potential, and how this steepness evolves as the field
moves along the potential.  For a given field potential they 
also allow the dynamics which follow from a more general potential 
to be expanded locally about the dynamics which follow from a
scaling potential with the same local steepness. The power 
spectrum which follows from the general potential can then also be written
in terms of the slow-roll parameters.

For the case at hand we would like to develop a similar expansion
scheme. 
However, for
the regime which we are considering in which $\bar{\epsilon} \gg 1$, it is clear that the
kinetic energy is of approximately the same magnitude as the potential energy, and
therefore the slow-roll approximation is inadequate. Indeed in this case
the field is evolving rapidly along a steep negative potential,
and we refer to the evolution as the `fast-roll' regime.
Our strategy will therefore be to determine other suitable small
parameters which characterise the steepness and curvature of the
potential, and which we will refer to as `fast-roll' parameters. The  
derived parameters we will arrive at are similar to those obtained in 
Ref. \cite{Gratton:2003pe}, where fast-roll parameters were required to parametrise
general potentials in the ekpyrotic scenario. This similarity is
natural since, as we have already
mentioned in the introduction, the evolution of the
super-inflationary scaling solution in LQC is dual to the ekpyrotic collapse.

The first step in accommodating a more general class of potentials is to allow 
$\bar{\epsilon}$ to become time dependent. From its definition it then follows that
\begin{equation}
\bar{\epsilon}' = - (2\bar{\epsilon})^{3/2}\,  \eta \sqrt{\frac{S}{D}} \phi' \,,
\end{equation}
where we have defined
\be
\eta \equiv 1-\frac{V_{,\phi\phi} V}{V_{,\phi}^2}
-\frac{1}{2} \, \frac{V}{V_{,\phi}} \, \left( \frac{D_{,\phi}}{D}-\frac{S_{,\phi}}{S}\right) \,,
\ee
which can also be written in terms of the background quantities as
\be
\eta \equiv 1-\frac{V_{,\phi\phi} V}{V_{,\phi}^2}
-\frac{1}{2} \, (n-r) \frac{V}{V_{,\phi}} \frac{a'}{a \, \phi'} \,.
\ee
Likewise, we can calculate $\eta'$ in terms of a third parameter $\xi^2$,
\be 
\eta' = - \sqrt{2\bar{\epsilon}} \, \xi^2 \sqrt{\frac{S}{D}} \phi' \,,
\ee
where 
\begin{eqnarray}
\xi^2 &\equiv&  \left[1+ \frac{V_{,\phi\phi\phi} V}{V_{,\phi\phi} V_{,\phi}} - 2 \frac{V_{,\phi\phi}V}{V_{,\phi}^2} \right]
\frac{V_{,\phi\phi}V}{V_{,\phi}^2} + \nonumber \\
&~& -\frac{1}{2} \left[1+ \frac{D_{,\phi\phi} \, V}{D_{,\phi}\, V_{,\phi}} - \frac{D_{,\phi} V}{D V_{,\phi}} - \frac{V_{,\phi\phi}V}{V_{,\phi}^2} \right] \frac{D_{,\phi} V}{D V_{,\phi}} + \nonumber \\
&~& + \frac{1}{2} \left[1+ \frac{S_{,\phi\phi} \, V}{S_{,\phi}\, V_{,\phi}} - \frac{S_{,\phi} V}{S V_{,\phi}} - \frac{V_{,\phi\phi}V}{V_{,\phi}^2} \right] \frac{S_{,\phi} V}{S V_{,\phi}} \,.
\end{eqnarray}
In particular, for the scaling solution where $\bar{\epsilon}$ is constant, one can verify that $\eta = \xi^2 = 0$ exactly.
Since we are considering potentials which are close to the form of a
scaling potential, we expect that there will be solutions to the equations of
motion of a form very similar to that given by Eqs.~(\ref{atau}), (\ref{defp}), (\ref{phiprime}) and (\ref{Vtau}), when $\bar{\epsilon}$ is slowly varying.
Assuming that Eqs.~(\ref{atau}), (\ref{defp}) and (\ref{phiprime}) are indeed good approximations, we have in general that
\be
\frac{d \ln \bar{\epsilon}}{d \ln a} \approx 4 \frac{\bar{\epsilon} \, \eta}{\alpha} \,, \hspace{1cm}
 \frac{d \ln \eta}{d \ln a} \approx 2 \frac{\bar{\epsilon} \, \xi^2}{\eta \, \alpha} \,.
 \ee
Imposing that $\bar{\epsilon}$ and $\eta$ are
 slowly varying, and since $\bar{\epsilon}$ is large in the
 regime which gives rise to scale invariance,
 requires $\eta$ and $\xi^2$ to be small, i.e. the potential must be
 nearly power law in form which is in agreement with our assumption.
We refer to $\eta$ and $\xi$ as the second and third fast-roll parameters,
and for convenience introduce the first fast-roll parameter as
$\epsilon = 1/2\bar{\epsilon}$,  where in terms of relative magnitude
we find $\epsilon \sim \eta$ and $\xi^2 \sim {\cal O}(\epsilon^2)$.
Using these relations it is possible to verify by substituting Eqs.~(\ref{atau}) -- (\ref{Vtau}) into the Friedmann and equation of motion that these solutions are indeed valid up to second order in fast-roll parameters for a general negative potential confirming the consistency of our analysis.

Then using Eqs.~(\ref{atau})-(\ref{Vtau}) to substitute for the respective 
quantities in the effective mass (\ref{meff}), and expanding to first order in the parameters $\epsilon$ and $\eta$, we obtain 
\be
\label{tilt}
{\Delta n}_u \approx 4\epsilon \left[1 - \frac{n}{12}\left(1+\frac{n}{6}-r\right)-\frac{r}{2}\right] -4\eta \,.
\ee
We note that although we have used the solution to Eq.~(\ref{w''}) which
is valid only when $m_{\rm eff} \tau$ is constant, the solution will remain sufficiently
accurate provided that $\epsilon$ does not vary
significantly as a given  $k$ mode crosses the horizon from the small wavelength to the
long wavelength regime. This is simply the condition that $\eta$ is
small, which we have assumed already. 
The spectral tilt can therefore be calculated at any given
scale by inserting into Eq.~(\ref{tilt}) the values that $\epsilon$ and $\eta$ take as this
scale crosses the horizon. In particular, to compare with observations
we need to consider the mode which corresponds to the largest scales on
the Cosmic Microwave Background (CMB).

Finally we also note that we could have included time derivatives of the
power $n$ and $r$. The generalisation to do so is straightforward but  
for simplicity the  computation we have presented does not include this possibility.

%%%%%%%%%%%%%%%%%%%%%%%y
\subsection{Stability of the fixed points}
%%%%%%%%%%%%%%%%%%%%%%%%

The analysis we have performed so far is intriguing.  It appears that
in the $a \ll a_*$ regime, the scaling solution which follows from a
steep negative potential can give rise to a scale-invariant power
spectrum of scalar field perturbations, and moreover we can
generalise the analysis to potentials which deviate from the scaling potential. 

However, there is another element which is involved in building a
convincing theory for the origin of scale-invariant perturbations.
That is, it would be highly desirable if the scaling solution was an
attractor, 
so that initial conditions not 
exactly on the solution would evolve towards it, and the solution's
stability against small local perturbations would be assured.

\begin{figure}[]
\includegraphics[width = 8.5cm]{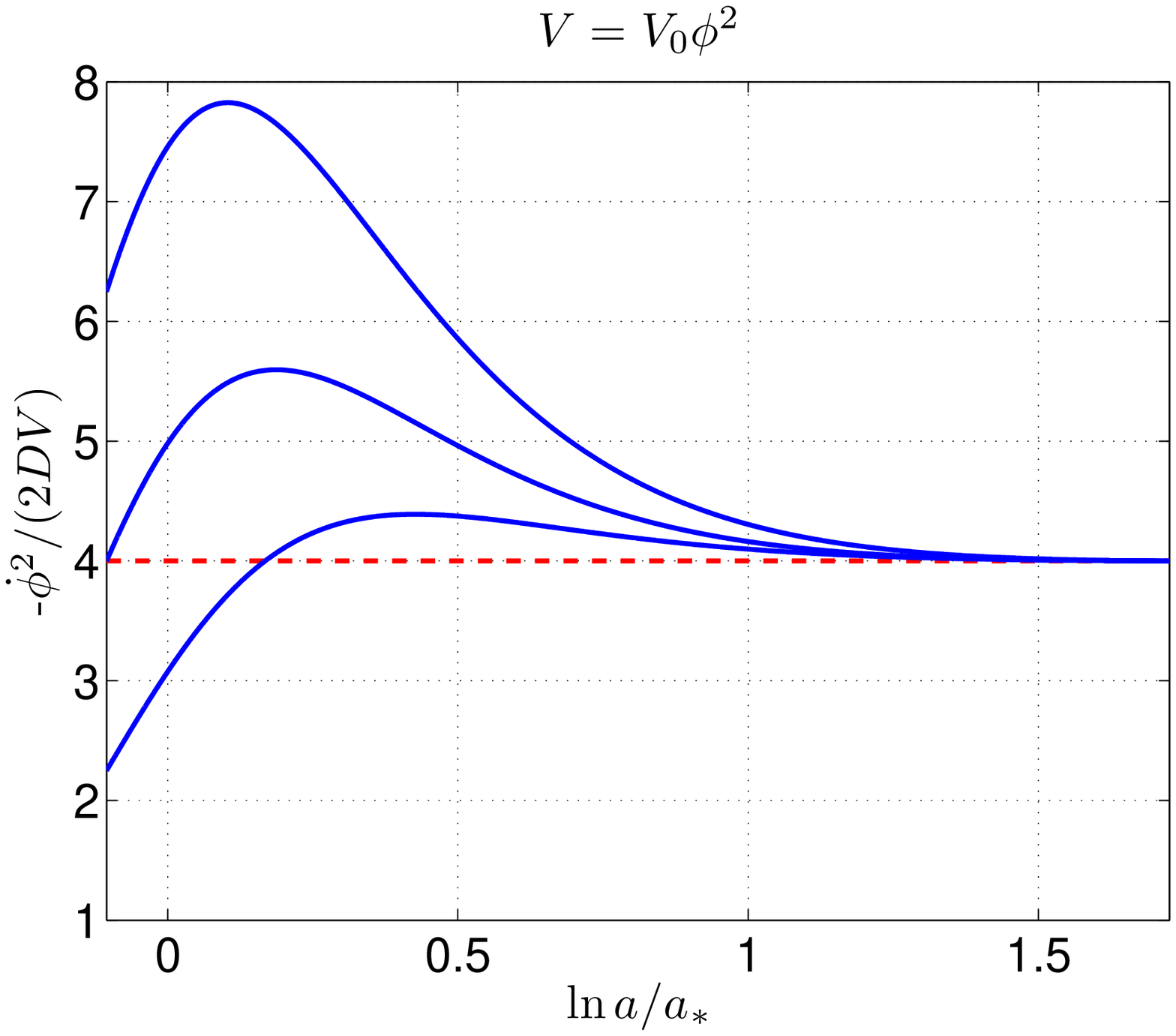}
\caption{The evolution of the ratio $-\dot{\phi}^2/2DV$ obtained by numerically solving the equations of motion for three different initial conditions (solid line). They approach the scaling solution given by $x^2/y^2$ where $x$ and $y$ are given by Eqs.~(\ref{FixedPoints2}) (dashed line). We used as parameters $V_0 = -10^{-20}$, $\phi_{\rm init} = 1$, in Planck units and $n=15$, $r = 3$ and $a_{\rm init} = 0.9 a_\star$, in a flat universe.}
\label{scaling1}
\end{figure}

To determine the stability of the scaling solution, we study the 
nature of the fixed points of Eqs.~(\ref{xN}) and (\ref{lambdaN}).  We
do this by linearising the equations about the fixed points and
determining the corresponding eigenvalues ($\omega$) in each case.
For the kinetic energy dominated solutions, valid for an arbitrary $\lambda$, 
in Eqs.~(\ref{FixedPoints1}) and (\ref{FixedPoints3}), we
find their respective eigenvalues to be 
\begin{eqnarray}
\omega_+ = 6 \alpha + \sqrt{6} \lambda \,, \hspace{1cm}  \omega_- = -\frac{1}{2} (n-r) \,, \\
\omega_+ = 6 \alpha - \sqrt{6} \lambda \,, \hspace{1cm}  \omega_- = -\frac{1}{2} (n-r) \,.
\end{eqnarray}
Since $n>r$ and $\alpha < 0$, the first fixed point is stable for $\lambda < -\sqrt{6} \alpha$ and the second for $\lambda > \sqrt{6}\alpha$.
Turning to the scaling solution, Eq.~(\ref{FixedPoints2}), we find
\begin{eqnarray}
\omega_\pm = -\frac{1}{4 \alpha} \left( \theta \pm \sqrt{ \theta^2 + 8\, \alpha\, (n-r) (\lambda^2 - 6 \alpha^2} \right) \,,
\end{eqnarray}
where $\theta = \alpha \, (6-r) - \lambda ^2 < 0$. The scaling solution is therefore stable whenever $\lambda^2 > 6 \alpha^2$ which coincides with the region of existence of this solution.
In Fig.~\ref{scaling1} we show the evolution of the ratio $-\dot{\phi}^2/2DV$ obtained by numerically solving the equations of motion.  We can see that it approaches the value given by $x^2/y^2$ where $x$ and $y$ are given by Eqs.~(\ref{FixedPoints2}). Tyically the evolution only reaches the attractor when $a > a_\star$ which is far outside the region where the appoximation $D \approx a^n$ is valid. We conclude that this solution must be extremely
fine tuned in order to deliver the dynamics and power spectrum as
described in the previous
subsections.

In the second part of this article we will be dealing with a
second possibility of 
obtaining a super-inflationary regime that also leads to a scale 
invariant power spectrum 
with the advantage that the stability of the 
scaling solution is no longer a dangerous issue.

%%%%%%%%%%%%%%%%%%%%%%%%
\section{Effective dynamics with quadratic corrections}
\label{rhosq}
%%%%%%%%%%%%%%%%%%%%%%%%

The second modification to classical dynamics which we consider follows from
considering that holonomies are the basic variables for quantisation
in LQC.  This modification gives rise to a Friedmann equation of the following form \cite{rho2}:
\be
\label{H^2_nonmin}
H^2 = \frac{1}{3} \, \rho \, \left(1- \frac{\rho}{2\sigma}\right) \,.
\ee
Once again we are assuming either a flat universe or that the curvature contribution can be safely neglected.
It is interesting that this form of the Friedmann equation is
identical to the form which arises in braneworld scenarios with a single
time-like extra dimension in the absence of
a black hole in the bulk 
spacetime \cite{Shtanov:2002mb}. In the braneworld case $\sigma$
represents the brane tension while for the LQC case $2\sigma$ represents the critical energy density arising from quantum geometry effects which leads to the scale factor undergoing a bounce as $\rho$ approaches it.

We are interested in high density regimes where $\rho$ approaches the bounding 
value of $2 \sigma$. In this case, the term within brackets tends to zero, and the
behaviour of the 
equations alters significantly compared with the classical
behaviour. Indeed in this regime we have $\dot{H}>0$ and
for an expanding universe super-inflation takes place.  
We will again consider a scalar field dominated universe, hence, $\rho = \dot{\phi}^2/2 + V(\phi)$. We stress that we are studying inverse volume and quadratic corrections separately, hence we do not include the $D$ and $S$ functions in the Friedmann equation and in the definition of energy density.
The scalar field equation of motion 
\be
\label{e.o.m_nonmin}
\ddot{\phi}+3H\dot{{\phi}}+V_{,\phi} = 0 \,,
\ee
is unchanged from the classical form.

%%%%%%%%%%%%%%%%%%%%%%%%
\subsection{Scaling dynamics}
%%%%%%%%%%%%%%%%%%%%%%%%

It was shown in Ref.~\cite{Copeland:2004qe} (also see
\cite{Singh:2006sg}) 
that
the effective equations (\ref{H^2_nonmin})-(\ref{e.o.m_nonmin}) also
allow a 
scaling solution in which the
kinetic and potential energy vary in proportion to one another. In
this case the potential must be of the form $V= V_0{\rm cosh}(\phi)$. 
However, while earlier works claimed otherwise, 
it can be shown that the scaling solution is not an
attractor. 
Moreover, the scaling solution which was found implies an evolution
for the scale factor 
which is not of a power law kind, and this means that it is unlikely to give
rise to a scale-invariant spectrum of perturbations. 
Given these difficulties, instead of focusing on the scaling solution
we will ask whether there is a form of the scalar
field potential which does result in a power 
law evolution of the scale factor. 
We are interested in the regime in which $\rho \approx 2 \sigma$, 
when super-inflation occurs, and where $H \approx 0$. Inserting the power-law ansatz, 
\be
\label{anz}
a(t) = (-t)^m \,,
\ee
where $m<0$, into the time derivative of the Hubble rate,
\be
\label{HDot}
\dot{H} = -\frac{\dot{\phi}^2}{2} \left( 1-\frac{\rho}{\sigma} \right)  \,,
\ee
we see that for $\rho \approx 2\sigma$ the kinetic energy of the field is $\dot{\phi}^2/2 \approx -m/t^2$ which upon integration gives
\be
\label{phiapprox}
\phi \approx \pm \sqrt{-2m} \, \ln t \,.
\ee
Using Eq.~(\ref{H^2_nonmin}) and expanding in $6H^2/\sigma \ll 1$ we have that
\be
\rho \approx 2\sigma -3H^2 \,,
\ee
and it follows from the definition of the energy density of the field and Eq.~(\ref{phiapprox}) that
\be
\label{Vapprox}
V \approx 2\sigma - U_0 e^{-\lambda\phi} \,,
\ee
where $U_0 = 3m^2-m > 0$ and $\lambda = 1/\sqrt{-2m}$.
It is evident that, in this regime,
scaling exists between $V-2\sigma$ and the
kinetic energy $\dot{\phi}^2/2$.  We now look for a more precise
description of the dynamics. 
The form of the potential (\ref{Vapprox}) motivates us to define the new variables:
\be
\label{nonmin_xy}
x \equiv \frac{\dot{\phi}}{\sqrt{4\sigma-2\rho}}, \hspace{1cm} y \equiv \frac{\sqrt{U}}{\sqrt{2\sigma-\rho}}\,~,
\ee
such that $\rho  \lesssim 2\sigma$ and $V(\phi) = 2\sigma -
U(\phi)$. In terms of a system of 
first order differential equations, the equation of motion of the scalar field now reads,
\begin{eqnarray}
\label{xNb}
x_{,N} &=& -3 x - \sqrt{\frac{3}{2}} \, \lambda \, y^2 - 3x^3	\\
\label{yNb}
y_{,N} &=& - \sqrt{\frac{3}{2}} \, \lambda x y - 3 x^2 y	\\
\label{lambdaNb}
\lambda_{,N} &=& -\sqrt{6} \, \lambda^2 (\Gamma-1) x + 3 x^2 \left(\frac{2\sigma}{\rho}-1\right) \sqrt{\frac{2\sigma}{\rho}}  \,~,
\end{eqnarray}
where $\lambda$ and $\Gamma$ are defined as
\be
\label{nonmin_lambda_Gamma}
\lambda \equiv -\frac{U_{,\phi}}{U} \sqrt{\frac{2\sigma}{\rho}}  \,,
\hspace{1cm}
\Gamma \equiv \frac{U \, U_{,\phi\phi}}{U_{,\phi}^2}\,~.
\ee
The variables $x$, and $y$ are also
related by the constraint condition
\be
\label{Constraint_2}
x^2 - y^2 = -1.
\ee

Considering the regime discussed above where 
$2\sigma/\rho \approx 1$, we can see $\lambda$ is a constant and by integrating
$\lambda$, we see that the $U$ part of
the scalar potential is given by
\be
U = U_0 \,e^{-\lambda \phi} \,,
\ee
as we were expecting from Eq.~(\ref{Vapprox}).

We consider the section of the phase space in which
$x<0$, $y>0$, and $\lambda>0$.
Taking  $\lambda$ to be constant, and substituting
the constraint equation (\ref{Constraint_2}) into Eq.~(\ref{xNb}),
results in an autonomous system with three fixed points. Two of them are non-physical solutions with
\be
\label{FixedPoints1b}
x = \pm i \,, \hspace{1cm} y = 0 \,,
\ee
and the third is a scaling solution with
\be
\label{FixedPoints2b}
x = -\frac{\lambda}{\sqrt{6}} \,, \hspace{1cm} y = \sqrt{1+ \frac{\lambda^2}{6}} \,.
\ee
The scaling solution is valid for all real values of $\lambda$.

As in the previous case, it is straightforward to show that as the universe evolves
according to this solution, the scale factor undergoes a power
law evolution
\be
\label{ataub}
a(\tau) = (-\tau)^p \,,
\ee
where
\be
\label{defpb}
p = -\frac{1}{\bar{\epsilon}+1} \,,
\ee
and $\bar{\epsilon}$ is here defined as $\bar{\epsilon} =
(U_{,\phi}/U)^2/2 \approx \lambda^2/2$.  This is of course what we
expect since we began by searching for such a solution using the
ansatz Eq.~(\ref{anz}).  
The time derivative of the field and the potential yields,
\begin{eqnarray}
\label{phiprimeb}
\phi' &=& \frac{\sqrt{2}\bar{\epsilon}}{\bar{\epsilon}+1} \, \frac{1}{\tau} \,, \\
\label{Vtaub}
V &=& 2\sigma - \frac{3+ \bar{\epsilon}}{(1+\bar{\epsilon})^2} \, \frac{1}{(a\tau)^2} \,.
\end{eqnarray}
We are now ready to compute the spectrum of the scalar field
perturbations produced by this power-law solution.

%%%%%%%%%%%%%%%%%%%%%%%%
\subsection{Power spectrum of the perturbed field}
%%%%%%%%%%%%%%%%%%%%%%%%

In this section we follow the same approach we took in the previous
analysis of the scalar field perturbations.
In terms of conformal time,
the perturbation equation for the scalar field, $\phi$, can be written as
\be
\delta \phi'' = -2 \frac{a'}{a} \delta \phi' + \left( \nabla^2 -a^2 V_{,\phi\phi} \right) \delta \phi\,
\ee  
which in turn can be written in terms of $u \equiv a \delta \phi$ as
\be
u'' + \left( -\nabla^2 +m_{\rm eff}^2 \right) u = 0\,~,
\ee
and the effective mass of the field $u$ is
\be
\label{meffb}
m_{\rm eff}^2 = \left( -\frac{a''}{a}+a^2 V_{,\phi\phi} \right).
\ee

Decomposing $u$ in Fourier modes $w_k$, that satisfy
\be
\label{w''b}
w_k''+\left( - k^2+m_{\rm eff}^2\right) \, w_k = 0 \,,
\ee
the power spectrum is then given by
\be
{\cal P}_u = \frac{k^3}{2\pi^2} |w_k|^2 \,.
\ee

The general solution to Eq.~(\ref{w''}) is
\begin{eqnarray}
\label{w_knormalisedb}
w_{k}(\tau) = \sqrt{\frac{\pi}{4}} \,  \left( d_{1}\sqrt{-\tau} \,
H_{|\nu|}^{(1)}(x)
 +  d_{2} \sqrt{-\tau} \,H_{|\nu|}^{(2)}(x) \right) \,,
\end{eqnarray}
where the subscript $|\nu|$ is
\begin{equation}
\label{nudefb}
\nu = - \frac{\sqrt{1-4 \, m_{\rm eff}^2 \, \tau^2}}{2}  \,,
\end{equation}
and $d_{1}$ and $d_{2}$ are constants subject to the condition $|d_{1}|^2-|d_{2}|^2=1$ and $H_{|\nu|}^{(1)}(x)$ and $H_{|\nu|}^{(2)}(x)$ are Hankel functions of the first and second kind, respectively.
For large wavelength modes, the power spectrum can be approximated by
\begin{equation}
\label{Pub}
{\cal P}_{u} \propto  k^{3-2|\nu|} \, (-\tau)^{1-2|\nu|} \,.
\end{equation}

By substituting Eqs.~(\ref{ataub}) -  (\ref{Vtaub}) into Eq.~(\ref{meffb}), we find
\be
m_{\rm eff}^2 \tau^2 = -2+3p+3p^2.
\ee
Comparing this case  to our previous results, we expect to have scale
invariance for  $p\to 0$. 
Once again, therefore, scale invariance occurs when 
the field $\phi$ is rolling down a steep potential and
the kinetic energy is not negligible, but
comparable to $V-2\sigma$. Hence the evolution should again be
understood as a fast-roll regime.

Clearly, scale invariance is also obtained for $p \to -1$ or $\bar{\epsilon} \ll 1$ which corresponds to the standard slow-roll regime that we are not concerned with for the purposes of this work.

In order to extend this analysis to general potentials, as we did for
the previous system of modified equations we studied,
we allow $\bar{\epsilon}$ to depend on 
conformal time such that
\be
\bar{\epsilon}' = - (2\bar{\epsilon})^{3/2} \eta \, \phi' \,,
\ee
which defines the fast-roll parameter $\eta$,
\be
\eta \equiv 1- \frac{U_{,\phi\phi} U}{U_{,\phi}^2} \,.
\ee
Similarly, $\eta'$ can be written in terms of the next order fast-roll parameter $\xi^2$ as
\be
\eta' = - \sqrt{2\bar{\epsilon}} \, \xi^2 \, \phi' \,,
\ee
with
\be
\xi^2 \equiv \left(1+ \frac{U_{,\phi\phi\phi}  U}{U_{,\phi\phi} U_{,\phi}} - 2 \frac{U_{,\phi\phi} U}{U_{,\phi}^2} \right) \frac{U_{,\phi\phi} U}{U_{,\phi}^2} \,.
\ee

For the scaling solution, it can be verified that both $\eta$ and
$\xi^2$ vanish. As in the previous case, we can use
Eqs.~(\ref{ataub}), (\ref{defpb}) and (\ref{phiprimeb}) as approximate solutions, and hence we have
\be
\frac{d \ln \bar{\epsilon}}{d \ln a} \approx 4 \epsilon \eta \,,
\hspace{1cm}
\frac{d \ln \eta}{d \ln a} \approx 2 \frac{\bar{\epsilon} \xi^2}{\eta} \,,
\ee
which means that for a large and slowly varying $\bar{\epsilon}$ the
parameter $\eta$ must be 
small, and  for a slowly varying $\eta$ the parameter $\xi^2$ must
also be small. Hence, the $U$ part of the scalar potential must be close to exponential.

Using now Eqs.~(\ref{ataub}) - (\ref{phiprimeb}) in
the expression for the effective mass of $u$ Eq.~(\ref{meffb}), where
$\bar{\epsilon}$ is now time dependent,  
and then using Eq.~(\ref{nudefb}), we find that
${\Delta n}_u \equiv 3-2|\nu|$ is to first order
\be
\label{delta2}
{\Delta n}_u = -4 (\epsilon-\eta)
\ee
where we have again defined a further fast-roll parameter by 
$\epsilon = 1/2\bar{\epsilon}$.
Again the assumption that
$m_{\rm eff} \tau$ is nearly constant as a given mode evolves outside the
cosmological horizon is valid. Hence the spectral tilt
for a given $k$ mode can be calculated by inserting the 
values the fast-roll parameters took as the mode crossed the horizon in
Eq.~(\ref{delta2}).
It is clear that the system we have investigated here 
results in a scale-invariant spectrum of scalar field perturbations for
$\epsilon \ll 1$ and $\eta \ll 1$. 

%%%%%%%%%%%%%%%%%%%%%%%%
\subsection{Stability of the fixed points}
\begin{figure}[]
\includegraphics[width = 8.5cm]{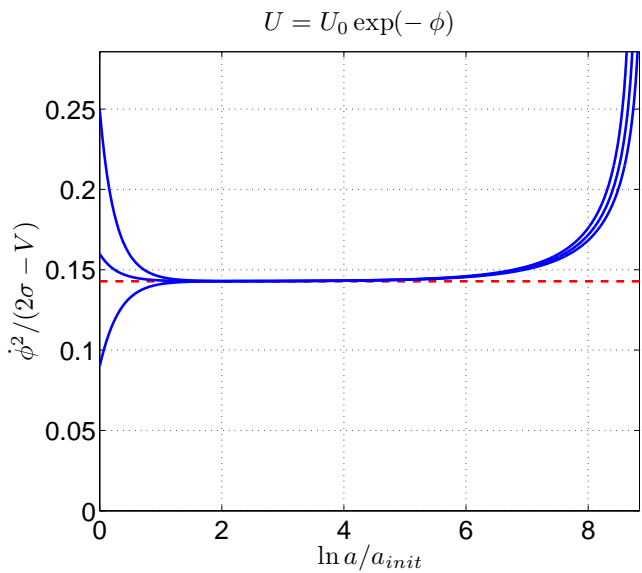}
\caption{The evolution of the ratio $\dot{\phi}^2/(2\sigma-V)$ obtained by numerically solving the equations of motion for three different initial conditions (solid lines). They approach the scaling solution given by $2x^2/y^2$ where $x$ and $y$ are given by Eqs.~(\ref{FixedPoints2b}) (dashed line). We used as parameters $U_0 = 10^{-2}$, $\phi_{\rm init} = 1$ and $\sigma = 0.41$ in Planck units, in a flat universe.}
\label{scaling2}
\end{figure}

%%%%%%%%%%%%%%%%%%%%%%%%
Linearising the system (\ref{xNb}), using Eq.~(\ref{Constraint_2}) about the fixed points
we find the following eigenvalues $\omega$: for the unphysical kinetic energy dominated solution, Eq.~(\ref{FixedPoints1b}),
\be
\omega = 6 \pm i \sqrt{6} \lambda \,,
\ee
and hence this solution is unstable.
While for the scaling solution, Eq.~(\ref{FixedPoints2b})
\be
\omega = -\frac{1}{2} (6+\lambda^2) \,,
\ee
and hence this point is a
{\it stable} attractor for {\em all} values of $\lambda$.
A numerical analysis of this system shown in Fig.~\ref{scaling2}
supports our analytical results presented here. The figure shows the evolution of the ratio $\dot{\phi}^2/(2\sigma-V)$ obtained by numerically solving the equations of motion for three different initial conditions. They approach the scaling solution given by $2x^2/y^2$ where $x$ and $y$ are given by Eqs.~(\ref{FixedPoints2b}). When 
$a \gg a_\star$ the quadratic corrections become negligeble and the numerical evolution diverges from this attractor.

%%%%%%%%%%%%%%%%%%%%%%%%%%%%%%%%%%%%%%%
\section{Number of $e$-folds}
\label{efolds} 

Before concluding, it is important to address the question of whether
a sufficient amount of super-inflation can occur in LQC to account for 
the largest scale perturbations observed on the CMB.  This in turn is equivalent
to asking whether the super-inflationary phases can solve the
cosmological horizon problem.  

It is clear that only a small number of $e$-folds of super-inflation can be
considered generic for either of the modified sets of evolution
equations we have studied \cite{Bojowald:2003mc}.  This might be considered
disappointing, since experience
from standard inflation suggests that approximately $60$ $e$-folds of
inflation are required for consistency with observations.  However, the
inflationary periods we have been studying are considerably different
to standard inflation and this has a dramatic effect. 

Solving the 
horizon problem is essentially the requirement that $aH$ 
grows sufficiently during an early stage of the universe's evolution. While in standard inflation this is
accomplished by $a$ changing rapidly as $H$ remains nearly
constant, in our case the converse appears to be possible, that $H$
increases sufficiently as $a$ remains nearly constant.  The number of 
$e$-folds usually only
refers to the change in $a$, and so we only expect a small number of e-folds
to be necessary in a super-inflationary phase, provided that $H$ changes
sufficiently.  To confirm the expectation that our super-inflationary
phases can indeed solve the horizon problem, let us now quantify our
qualitative arguments.

During super-inflation, perturbation modes exit the cosmological
horizon, and once super-inflation ends, modes start to re-enter.  
Let us consider
a perturbation mode with wavenumber $k$, such that $k$
exited the cosmological horizon $N(k)$ $e$-folds before the end of the
super-inflationary phase, and re-entered sometime later. The mode
re-entering the horizon 
today, $k_*$, must satisfy $k_*=a_0 H_0$, where
subscript $0$ indicates quantities at the present epoch. Comparing
this with the generic $k$ mode we have:
\be
\frac{k}{a_0 H_0}=\frac{a_k H_k}{a_0 H_0} = \frac{a_k H_k}{a_{\rm
    end} H_{\rm end}} \frac{a_{\rm end}}{a_{\rm reh}}\frac{a_{\rm reh}}{a_{\rm
    eq}}\frac{a_{\rm eq}}{a_{0}}\frac{H_{\rm end}}{H_0}\,~,
\ee
where subscript `end' labels quantities at the end of
inflation, `reh' at reheating, and `eq' at matter radiation
equality.  
Then employing the known
evolution of the universe from reheating to the present day, together with
the measured value of the Hubble rate at the present epoch, $H_0$, and for
simplicity assuming that the universe behaves as if it is matter
dominated between the end of inflation and reheating,
we find:
\begin{eqnarray}
\label{scales}
\ln \left[\frac{k}{a_{0} H_0} \right] &\approx&  68 + \ln \left[\frac{a_k H_k}{a_{\rm end} H_{\rm
    end}} \right]  - \frac{1}{2}  \ln \left[\frac{\mpl}{H_{\rm
    end}} \right] \nonumber \\  
    &~& - \frac{1}{3}\ln
\left[\frac{\rho_{\rm end}}{\rho_{\rm reh}} \right ]^{1/4} \,.
\end{eqnarray}
The energy scale at the end of inflation must be determined by requiring that the magnitude of the curvature perturbation accounts for the temperature anisotropies in the CMB. Since we have only worked with the scalar field perturbation this is so far undetermined in our model and we therefore take $H_{\rm end}$ to be the highest possible scale, i.e. $H_{\rm end} = \mpl$. A lower scale would lead to fewer required $e$-folds. Further considering $k = k_*$ and assuming instant reheating, we find that
\be
\label{scales2}
\ln \left(\frac{a_{\rm end} H_{\rm
    end}}{a_{k_*} H_{k_*}} \right ) \approx 68 \,.
\ee

In order to determine the number of $e$-folds of super-inflation
required, we now consider the cases in which the scale factor is
undergoing pure power-law behaviour, $a\propto (-\tau)^p$.  Using this
together with Eq.~(\ref{scales2}), we find
\be
\ln \left(\frac{\tau_{\rm end}}{\tau_{k_*}}\right )^{-1} \approx 68 \,,
\ee
and in turn 
\be
\label{e-folds}
N(k_*)=\ln \left (\frac{a_{\rm end}}{a_{k_*}}\right ) = -68 \, p \,.
\ee
Recalling that $p$ must be small and negative for scale invariance, we
see that only a small number of $e$-folds are required. 
Although considering behaviour which deviates from pure power law behaviour
will alter Eq.~(\ref{e-folds}), it is clear that the conclusion of only a
small number of e-folds being necessary will remain valid.

%%%%%%%%%%%%%%%%%%%%%%%%
\section{Discussion}
\label{discussion}
%%%%%%%%%%%%%%%%%%%%%%%%

In this paper we have investigated the nature of super-inflation in Loop Quantum Cosmology. 
Considering two specific examples which lead to modifications of the Friedmann equation, one 
where the modification is due to the presence of inverse volume corrections, and the second 
where it  is induced by the use of holonomies as the basic variable in the quantisation of Loop 
Quantum Gravity, we have demonstrated explicitly in both cases the existence of super-inflation 
solutions defined by $\dot{H}>0$ where $H$ is the Hubble parameter. Through the use of phase 
plane analysis we have been able to discuss the nature of the attractor solutions and their stability 
in both cases. Further we determined the scalar perturbations arising in these models, and 
through the introduction of fast roll parameters, showed that a class of potentials exhibit near 
scale invariant spectra. To be more specific we have shown that for the case with inverse volume 
corrections, if we concentrate on the regime $D \propto a^n$,  three  solutions exist, two of them 
corresponding to kinetic energy dominated solutions and one to a super-inflationary scaling 
solution where the kinetic energy scales in proportion to the potential energy. In this last case, 
when the potential is negative, scaling occurs when it is polynomial in form, and the perturbed 
scalar field equations for this potential are scale invariant (as first demonstrated in \cite{Mulryne:2006cz}). 
However, we have been able to go further. By introducing fast-roll parameters $\bar{\epsilon}$ 
and $\eta$ in an analogous manner to the way introduced for the ekpyrotic scenario \cite{Gratton:2003pe}, 
we were able to extend our calculation of the scalar power spectrum beyond the case of the polynomial 
scaling potential. In other words we can determine the degree to which scale invariance is broken in 
terms of the fast roll parameters. Through the stability analysis of these solutions we have seen 
that the scaling solution which gives rise to the scale invariant spectrum is a stable attractor. In general, however, the attractor is reached when $a \gg a_\star$ when $D \approx 1$ and therefore we concluded that in only a limited range of parameters we have an analytical understanding of the dynamics of the system in the semi-classical phase $a \ll a_\star$.

We considered a second set of corrections to the Friedmann equation, arising from the 
quantisation procedure in LQG where a $\rho^2$ term in the Friedmann equation, analogous 
to that found in particular braneworld models \cite{Shtanov:2002mb}, is present. In this case 
the scaling solution found in \cite{Copeland:2004qe}, which involves a $\cosh(\phi)$ potential, 
is not an attractor (in contradiction to the claims in Refs.~\cite{Singh:2006sg,Copeland:2004qe})  
and we believe it is unlikely to lead to a scale invariant spectrum. In order to obtain the solution 
with a scale invariant spectrum we decided to take another route, namely search for the 
potential with the correct time evolution in the scale factor which would lead to scale invariance. 
In these models super-inflation occurs just after the bounce and using this fact we quickly arrive 
at the intriguing result that the corresponding potential is an uplifted negative exponential 
potential, and in this case the scaling solution is stable. Scale invariance easily follows as 
does the generalisation to include a new set of fast roll parameters, hence obtaining potentials 
which will yield small deviations from scale invariance.

In both cases, we note that we do not expect the scalar potential to remain of the form we have used to give scale invariance throughout the entire evolution of the universe as a negative value of the potential after the end of super-inflation may lead to a subsequent recollapse. The full details of the exit from super-inflation, the subsequent form of the potential, as well as the 
transition to the radiation epoch, are beyond the scope of the present work.
 
We emphasize that we only dealt with the evolution of the universe in the expanding phase, however, for closed models, the universe may have gone through a bounce from a collapsing phase into the super-inflationary evolution. It is natural to question whether the bounce is symmetric or not, i.e. if the period of super-inflation is nothing more than the counterpart of an identical deflationary period and an epoch of standard inflation is still necessary. Though this is a possible situation, we stress that the scaling solutions produced by the potentials we considered here are the attractors only during the expanding epoch, hence, the evolution through the bounce is indeed expected to be asymmetric.  

Finally, we have discussed the horizon problem in this set up. We found that for the class of potentials studied that lead to scale invariance, the required number of $e$-folds of super-inflation 
is of only a few, more specifically, $N \approx -68p$ where $-1 \ll p < 0$ defines the time dependence 
of the scale factor of the universe, $a = (- \tau)^p$.  A previous 
criticism of infation within the inverse volume corrections approach has been that $a_\star$ must be unphysically large to give rise to $60$ $e$-folds \cite{Coule:2004qf}. We have seen in this work, however,  that only a small number of $e$-folds of super-inflation are required, which suggests that this bound on $a_\star$ can be evaded. We also note that previous studies found a small probability of sufficient standard inflation in the context of LQC \cite{Germani:2007rt}. However, their conclusions do not apply directly to our model of super-inflation since, as we have discussed, only a small number of $e$-folds are required and moreover, in their study, the Hubble rate was assumed to be constant.

In our calculation we have neglected  metric perturbations, instead we have only considered 
scalar perturbations and been able to find scale invariance through the scalar field perturbations.  
Given the similarity mentioned throughout the paper between our LQC results and the ekpyrotic 
mechanism, it seems likely that when metric perturbations are included in the LQC calculation, 
we will find that the scale invariant spectrum occurs in the Newtonian potential and the scalar field perturbation, 
but not in the curvature perturbation, just like the ekpyrotic case (for example see \cite{Gratton:2003pe}). 
If this turns out to be the case, it would  be interesting to consider whether the inclusion of a second 
scalar field in LQC, as in the `new ekpyrotic' mechanism, would allow  the scale invariant spectrum 
to be transferred from the scalar field perturbation into the curvature perturbation via an entropy 
perturbation \cite{Lehners:2007ac,Buchbinder:2007ad,Creminelli:2007aq}.

\begin{acknowledgments}
DJM is supported by the Centre for Theoretical Cosmology, Cambridge, NJN by STFC and
MS by a University of Nottingham bursary. The authors thank Roy Maartens for discussions 
and Parampreet Singh for helpful comments on the manuscript.
\end{acknowledgments}

\end{document}